# Activity Waves in Condensed Phases of Quincke Rollers

Meng Fei Zhang,[a] Bao Ying Fan,[a] Zeng Tao Liu,[a] Tian Hui Zhang[a]

[a] *Center for Soft Condensed Matter Physics and Interdisciplinary Research & School of Physical Science and Technology, Soochow University, Suzhou, 215006, P. R. China*

Wave-exciting is a universal phenomenon in physical and biological excitable systems. Here we show that colloidal systems of Quincke rollers which are driven periodically can condense into active liquids and active crystals, in which waves can be excited. In active liquids, the waves propagate antiparallel to local density gradients via the splitting of dense bands, and cross over each other in collision as sound waves do. The waves in active crystals have a sharp front like that of shock waves, and propagate parallel to local density gradients. The shock waves annihilate or converge as they collide. Detailed investigations on microscopic dynamics reveal that in sound waves, the dynamics of rollers is dominated by electrostatic repulsions; in shock waves, the dynamics is encoded with a density-dependent collective memory. These findings demonstrate a realization of excitable colloidal systems with tunable dynamics. This is of great interests in exploring the principles of self-organization and the fabrication of active functional materials.

## Introduction

In active matters, energy is transmuted to mechanical work at the single-particle level(1, 2). A rich of nonequilibrium patterns and collective motions have been realized and studied in active matters(3-5), ranging from bacteria(6), actin filaments(7), to fish schools(8) and bird flocks(9). To explore the general principles underlying collective motions, various active colloids have been developed and employed as model systems(10). In active colloidal systems, microscopic dynamics can be followed and quantified at the single-particle level. Most importantly, the interaction between colloidal particles, and thus the particle-level dynamics can be tuned (11, 12).

In active colloids, large-scale collective motions can emerge spontaneously by phase separation (13-19), giving rise to the coexisting of an ordered dense phase and a disordered gas phase. The dense ordered phase, in which all units move with a common velocity, exhibits a polar order (13, 16). Upon increasing the overall density, the scale of the polar phase increases correspondingly(5, 18). As the overall density is high enough, disordered jamming occurs (20, 21). Theoretically, however, it is supposed that if the interaction between active units dominate over the propulsion, condensed active phases, active crystals, can form (22-24). Active crystals exhibit a nontrivial dynamical response, and have potential applications in fabricating a new class of active matter with unusual rheological, phononic, and possibly, also photonic properties (25-27). As experimental models, active crystals also serve a type of intriguing systems to explore the dynamics of nonequilibrium systems (28, 29). Experimentally, however, the dynamics of active particles is generally dominated by propulsion (13, 15), and so far, it is still a challenge to achieve uniform condensed active systems experimentally in active colloids.

Here, in this study, we show that as Quincke rollers are subjected to a square-wave electric (SWE), they can condense into continuum active fluids or active solids at high frequencies as their dynamics is dominated by attraction. In the condensed active phases, density fluctuations above a critical magnitude develop into dense bands which travel like waves. In active liquids, the traveling bands cross over each other in collision. Whereas, the traveling bands in active crystal converge or annihilate in collision. Observations on the microscopic dynamics reveal that the formation and propagation of sound waves originate from the synchronized reset of propulsion and density-dependent dipolar repulsion, which breaks the dynamic symmetry of Quincke rotation and leads to directional particle flows antiparallel to local density gradients. The shock waves in active crystals arise from the coupling between propulsion and a density-dependent collective memory.

## Experiment Methods

Quincke rollers driven by electric field have been employed to explore the mechanisms of collective motions (13-15). In this study, polystyrene particles of a diameter of $9.9\ \mu m$ are dispersed in a mixture of AOT/hexadecane, and then the suspension is sealed in a cell constructed by two ITO-coated glass slides (Fig.1a). As a uniform electric field $\vec{E}$ above a critical magnitude $E_c$ is applied, the colloidal particles are driven to rotate and move (13, 30) (Materials and Methods). At subcritical field $E < E_c$, Quincke particles aggregate and form crystalline structures due to the long range attraction induced by electrohydrodynamic(EHD) flows (31). At $E > E_c$, isolated particles move at an equilibrium speed $v_e \sim \sqrt{(E/E_c)^2 - 1}$ (32), and the EHD-induced attraction can no longer bond and confine the particles. As the global density is above a critical value, large-scale collective motions emerge (13, 15). In experiments, however, as the global area fraction is above 0.50, it was found that collective motion is suppressed and jammed particles aggregate to form a dense disordered phase (21). In this study, a global area fraction $\varphi \sim 0.43$ is employed. This is much higher than that employed in previous studies (31, 33-35) but lower than the jamming density.

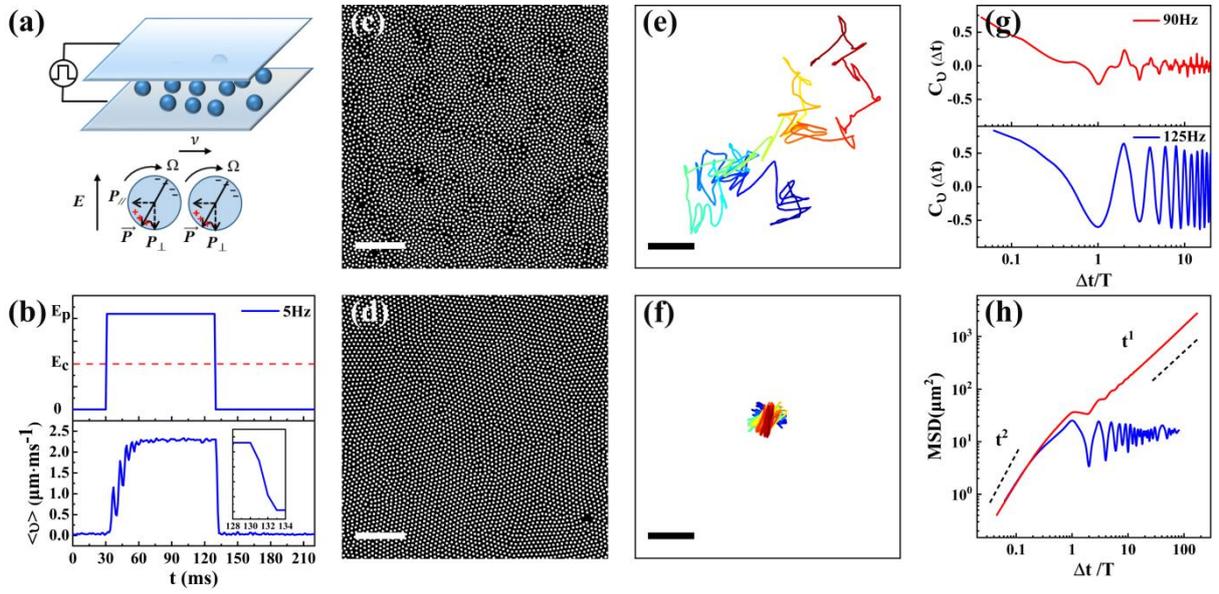

**Figure 1** Experiment setup and active condensed phases. a. Quincke rollers driven by electric field. b. The speed of isolated Quincke rollers as a function of time as the system is subjected to a square-wave electric field. c. Active liquid formed at $E_p = 2.1 E_c, f = 90 Hz$. The area fraction is $\varphi = 0.53$. d. Active crystal formed at $E_p = 2.1 E_c, f = 125 Hz$. The area fraction is $\varphi = 0.50$. e. A typical trajectory of liquid particles in 1.0s. f: The trajectory of one crystal particle in 1.0s. g. Velocity-velocity autocorrelation in active liquid and active crystal. h. Mean squared displacements of liquid particles and crystal particles. Scale bar in c and d:150μm. Scale bar in e and f: 2μm.

As driven by an oscillating electric field, Quincke rollers exhibit a variety of dynamic behaviors, and form intriguing patterns as the density fluctuations are periodically modified by directed particle flows (34-37). In this study, a square-wave electric field (SWE) is employed to drive Quincke rotation. The magnitude of SWE oscillates periodically between zero and $E_p (> E_c)$ (Fig. 1b, up). The durations for on-period (zero) and off-period ($E_p$) are equally set as $T/2$ in one cycle. $T$ is the period of SWE. Driven by SWE, the translational speed $v$ of Quincke rollers oscillates periodically between a running state ($v > 0$) and a rest state ($v = 0$) (Fig. 1b, bottom). In the on-period of $E_p$, Quincke rollers experience an acceleration of tens of microseconds (~20ms) before they reach an equilibrium speed $v_e$. As $T/2$ is smaller than the acceleration time, the particles cannot reach their equilibrium speed $v_e$ at all time. The maximum speed $v_p (< v_e)$ becomes frequency-dependent and decreases upon the increase of frequency (38). Meanwhile, as $E$ drops from $E_p$ to zero, the speed does not relax to zero immediately but experiences a deceleration (~3ms). In practice, the acceleration time is sensitive to the conductivity of solvent and exhibit a large fluctuation in experiments. In contrast, the deceleration time is stable and independent of $E_p$ (38).

**Results**

In experiments, as the maximum speed $v_p$ decreases with frequency, the EHD-induced attractions become dominant increasingly in determining the dynamics of Quincke particles. At high frequencies ($> 60Hz$), as the kinetic energy of particles cannot break up the bonding of the EHD-induced attraction, Quincke particles began to aggregate and form condensed phases. Because of the condensation, an empty region forms between the cell boundary and the condensed phases (SFig.1a and 1b).

Two distinct condensed phases, active liquid and active crystal, are realized in our experiments (Movie 1 and Movie 2). In active liquids (Fig.1c), the distribution of particles is homogeneous and there is no long-ranged pair correlation (SFig.1c). Both the trajectories (Fig.1e) and the Velocity-velocity auto correlation (Materials and Methods) show that the motions of particles in different cycles are not correlated (Fig.1g, up). In active crystals (Fig.1d), the pair correlation exhibits sharp peaks of hexagonal symmetry (SFig.1d). Nevertheless, because of the defects as can be seen in Fig.1d, the long-ranged correlation is destroyed. The trajectory (Fig.1f) shows that the motions of particles in active crystals are caged as in normal crystals. The Velocity-velocity autocorrelation suggests that the motions of particles in active crystals are periodic and oscillating locally (Fig.1g, bottom). In consistent, the mean squared displacements (MSD) of particles in the active crystal increases like ballistic motions at the short time scale ($< T$), and saturate with oscillation on a plateau at long time scale ($>T$). In active liquids, however, the particles exhibit a short-time ballistic motion as well but their MSD increases linearly at long time scale(Fig.1h), being consistent with typical Brownian motions. Nevertheless, it is noticed that the mean displacements in active liquid are only a few diameters of particle in 100 cycles. It follows that the diffusion of liquid particles is limited at the time scale of observations although a global diffusion is permitted.

Distinct from normal colloidal systems, the active condensed phases are excitable, in which waves can emerge and propagate (Movie 3 and Movie 4). The waves originate from local density fluctuations which create local denser regions (Fig.2a and c). The denser regions then grow up and develop into bands which propagate like waves (Fig.2b and d). As a common feature, the speed of individual particles in the bands is much higher than that in the bulk (Fig.2e-h). However, the bands in active liquid and active crystal are characterized by distinct density profiles, and propagate by different mechanisms.

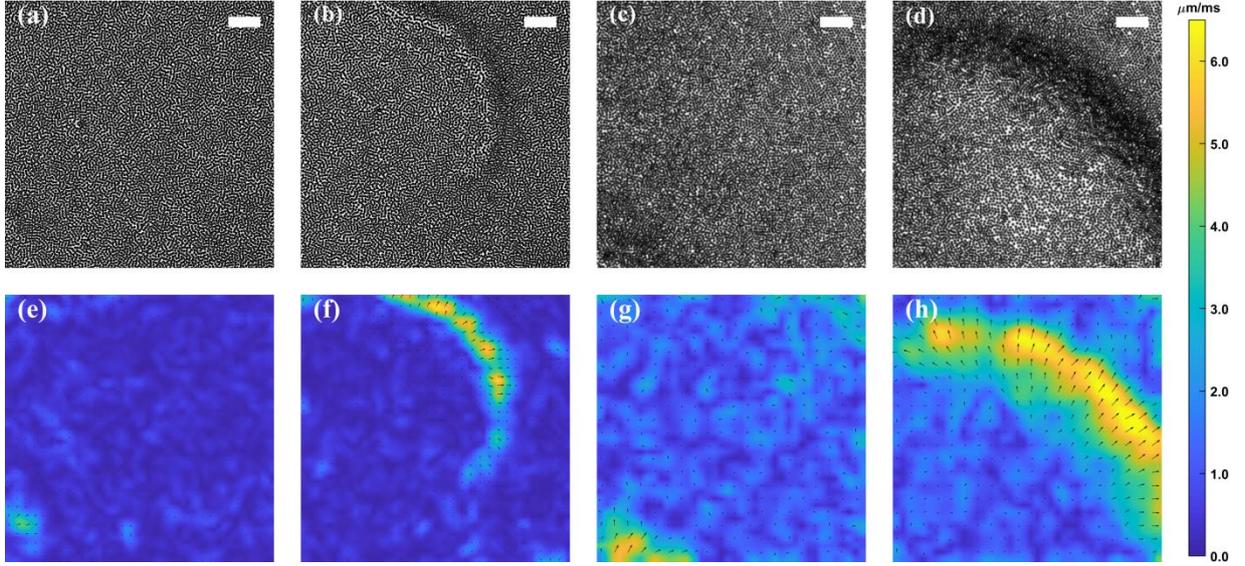

**Figure 2** Emergence of waves. a-b: Snapshots in the emergence of wave in active liquid at $E_p = 2.2 E_c$ and $f = 95$Hz. e and f are the corresponding velocity field created by PIV. c-d: Snapshots in the emergence of wave in active crystal at $E_p = 2.3 E_c$ and $f = 100$Hz. g-h: The corresponding velocity fields of c and d. Scale bar in a-d: 150μm.

The bands in active liquids are separated from the bulk at the rear by a low-density (area fraction) belt, giving rise to a sharp change in density (Fig.3a and Fig.3e up). In the front of bands, the density decreases gradually down to the bulk value in the direction of propagation. As $E$ is reset to $E_p$, particles in bands are activated but move with two opposite directions (Movie 5, Fig.3b-c). In front of the density peak, particles are propelled to move in the direction of propagation, giving rise to a positive $<v_x/|v|>$ (Fig.3f-h middle). Particles behind the density peak run backward as indicated by the negative $<v_x/|v|>$. As a result, the band splits into two parts. Outside the dense bands, however, particles are activated with random directions and give rise to a zero value of $<v_x/|v|>$. The particles moving backward fill the low-density belt at the rear while a new low-density belt is created at the position where the band splits (Fig.3h). Simultaneously, particles moving forward create a new density peak. In the following cycle, the particles moving forward will be reflected back by the splitting. Therefore, the displacements of individual particles included in waves are limited (SFig.2) whereas the bands propagate globally in the liquid. By following the positions of density peak, the wave speed in active liquids is measured as $\sim 6.93~\mu m/ms$ (Fig.3h up) which is much higher than the maximum speed of particles in bands (Fig. 3g bottom).

In active crystals, the density of band peaks in the front of band, and forms a sharp interface between the band and the bulk (Fig. 4a and 4e up). At the rear, the density decreases gradually to the bulk. In active crystals, particles in dense bands are activated and move collectively with a common direction (Movie 6). No splitting occurs at the density peak (Fig.4b-d and f-h). The speed in the band is not uniform (Fig.4g bottom):

Behind the density peak, the speed decreases with density such that tail particles drop gradually apart from the band as the bands propagate forward. Nevertheless, as the bands move forward, particles in front of the bands are collected. The balance between the collecting and the dropping leads to a stable width of bands. By following the positions of density peaks in different cycles (Fig.4h up), the wave speed is measured as $\sim 4.74~\mu m/ms$, which is comparable to that of individual particles in the bulk but is smaller than the maximum speed of particles in bands. As particles are involved in shock waves, their motions become well directed and exhibit a long distance displacement up to tens of diameter (SFig.3). Before they drop off the wave band, they move with the wave for tens of cycles. Nevertheless, the directional displacement of wave particles is not constant in experiments. It depends on the width of shock waves, which varies from case to case. The passing of wave destroys the local crystalline structure, but it will recover spontaneously after the passing (Movie 6).

In active liquids, the propagation of dense bands is coupled with the splitting. The splitting of dense clusters has been observed in Quincke systems subjected to low-frequency SWE (34-37). The mechanism is that as driven by SWE, the propulsion and the dipolar repulsion between Quincke rollers are switched on and off simultaneously in each cycle. The synchronized resetting of propulsion and repulsion breaks the dynamic symmetry of Quincke rotation and leads to the directed motion antiparallel to the density gradient: The motion is directed from dense regions where the repulsion is stronger to sparse regions. In active solids, however, no splitting occurs although the caged motion of particles indicates that the interaction between particles is significant. More intriguingly, it is noticed that in active crystals, the speed of particles in the dense bands changes periodically between zero and a peak value as well as that in the bulk, suggesting that the motion of particles in dense bands is not persistent but intermittent.

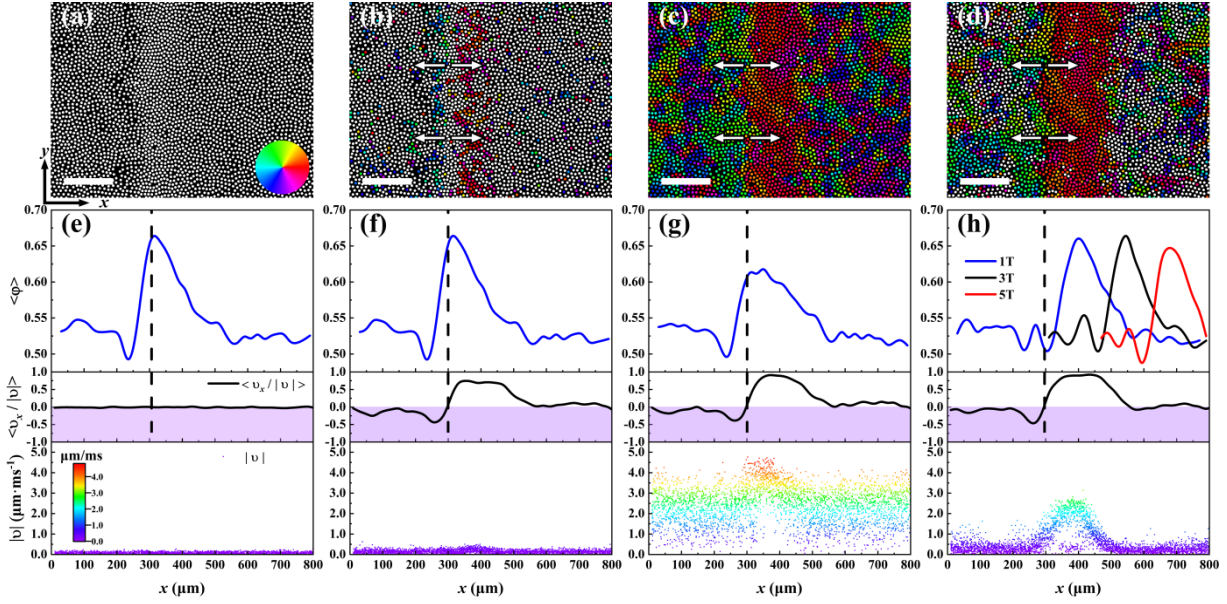

**Figure 3** Evolution of density and speed in one cycle of a sound wave in the active liquid at $E_p = 2.5E_c$ and $f = 100$Hz. a-d: Snapshots with particles colored by the direction of velocity. a: At the static state. b: At the starting of $E_p$ (0ms). c: 2ms after the starting of $E_p$. d: At the end of $E_p$(5ms). e-h: The corresponding density and speed distribution in the wave. $x$-axis is parallel to the direction of propagation. Up panel of e-h: The density is averaged over all particles with the same $x$. The positions of density peak in the wave at the end of different cycles are used to estimate the speed of wave. The region with a density higher than 0.55 is identified as the band of wave. The width of wave is around 300 μm which is universal for all waves in active liquids. Middle panel of e-h: The velocity component $v_x$ in $x$-direction is normalized by the magnitude of speed and then averaged over all particles with the same $x$. Bottom panel: Speed of particles. Each point represents one particle. The points are colored by speed and located by their $x$. Scale bar in a-d: 150μm.

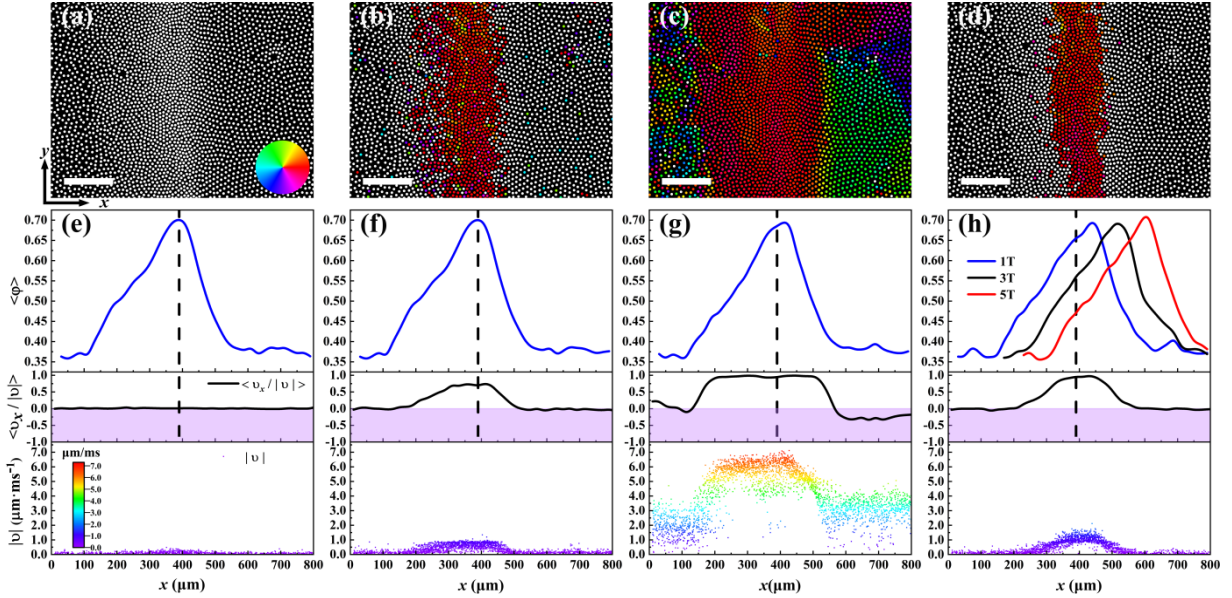

**Figure 4** Evolution of density and speed in one cycle of shock waves in active crystal at $E_p = 2.5E_c$ and $f = 120$Hz. a-d: Snapshots with particles colored by the direction of velocity. a: At the static state. b: At the starting of $E_p$ (0ms). c: 2.4ms after the starting of $E_p$. d: At the end of $E_p$(4.2ms). e-h: The corresponding density and speed distribution in the wave. Up panel: The density distribution in the wave. Middle panel: Normalized $v_x$ which is averaged over all particles with the same $x$. The direction of $v_x$ in wave is highly aligned, giving rise to a value of $\langle v_x/|v|\rangle$ close to 1 while it is positively or negatively close to zero out of the wave. Bottom panel: Speed of particles. Each point represents one particle. Scale bar in a-d: 150μm. The width of the presented wave is around 450 μm. The width of wave in active crystal varies from case to case.

However, the persistent direction of velocity in the bands suggests that the wave particles have a memory of the direction of propulsion. This memory is absent in the bulk where the density is relatively lower. The density-dependent memory of Quincke particles has been observed and reported recently (38). The mechanism is that as the relaxation time of charge is longer than the off-period of $E$, a memory of the dipolar and then the

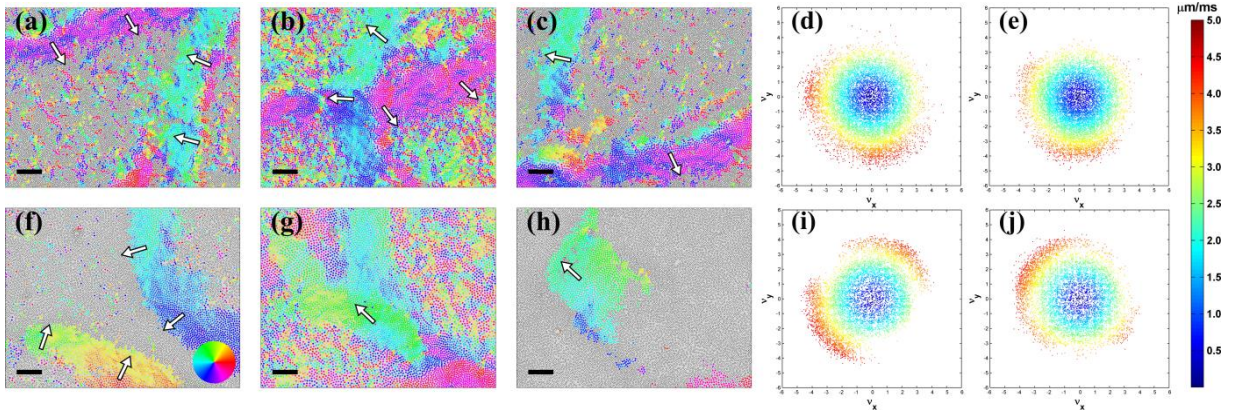

**Figure 5** a-c. Collision of sound waves. at $E_p = 2.25E_c$ and $f = 85$Hz. a. Before collision. b. In collision. c. After collision. d-e: The velocity distribution in the sound waves before and after collision. The particles are colored in the direction of their velocity, and each point represents one particle. f-h: The Snapshots of two colliding shock waves before collision, in collision and after collision at $E_p = 2.25E_c$ and $f = 95$Hz. i-j: The velocity components in the shock waves before and after collision. Scale bar: 150μm.

direction of propulsion emerges. Nevertheless, because of the dipolar interactions between Quincke rollers, the charge relaxation is slower in dense regions than that in sparse regions, giving rise to a density-dependent collective memory of the propulsion. In active crystals, the local density in wave bands is higher than that in the bulk such that the charge relaxation in wave is slower. Such that, a memory of propulsion emerges in wave while it is absent in the bulk.

To verify the role of density-dependent collective memory in the propagation of shock waves, a band traveling in the active crystal is frozen by switching off the electric field, and maintained for more than half an hour to assure that the dipolar moments in all particles are completely relaxed, and the memory of propulsion is cleaned in the system. After that, the SWE is switched on again. It is found that the frozen band splits into two bands immediately in the first cycle (Movie 7), suggesting that as the memory is removed, the dynamics is dominated by repulsion. However, in the following cycles, the resulting two bands propagate oppositely as two independent shock waves, and no splitting occurs anymore, suggesting that the memory is rebuilt after the first cycle. The splitting and the reemergence of memory confirm that the propagation of bands in active crystals results from the density-dependent collective memory.

As two traveling bands meet, a high-density collision line between the wave fronts forms. In active liquids, the reactivation of particles is coupled with electrostatic repulsion such that the particles are reflected back from the collision line (Fig.5a-c and Movie 8). As a result, the two bands are reflected backward with exchanged propagation directions. In effect, it looks like that the two bands get cross each other and propagate independently. The velocity components before (Fig.5d) and after (Fig.5e) the collision exhibit no big difference. In active solids, particles with memory of propulsion tend to persist in their own directions. Therefore, in the regions where waves collide with opposite directions, they counteract each other and leads to the annihilation of waves (Fig.5f-h and Movie 9). In regions where the waves collide with an acute angle, they converge and propagate in a new direction (Fig.5i-j), giving rise to new velocity components.

Five distinct phases are observed in this system. Figure 6 presents the phase diagram in the plane of $E_p$ and $f$. At low frequencies, dynamic stripe patterns form (Movie 10 and SFig.4a). Similar patterns have been reported in our previous studies (34). Therein a much lower area fraction (0.3) is employed, and two distinct patterns, square patterns and stripe patterns, can be realized. Here, to achieve condensed phases, a much higher global area fraction 0.43 is used, and only stripe patterns can emerge. The splitting of stripes in each cycle indicates that the dynamics in stripe patterns is dominated by dipolar repulsions. To form dynamic patterns, long-distance diffusion is necessary to form large-scale density fluctuations. Upon the increase of frequency, the persistent time of $E_p$ and thus the displacements of particles in each cycle decrease. The effective activity of particles is reduced as well. Upon the increasing of frequency, the low-activity particles begin to aggregate due to the EHD-induced attraction, giving rise to the active liquids in which sound waves can be excited. As the displacements become comparable to the mean distance between particles, active crystals form. The active crystals have two distinct sub-regions. In the low-frequency region, the active crystals are excitable while in the high-frequency region, no wave can be excited. As frequency is further increased, all particles have a memory of propulsion and velocity such that particles begin to move persistently and large-scale flocking emerges (Movie 11 and SFig. 4b).

**Discussion**

In previous studies, activity waves have been observed in sparse Quincke systems which are driven by a constant subcritical $E$ ($<E_c$) (31). The activity waves therein arise from the density-enhanced propulsion, and the particles involved in activity waves move persistently till they drop off from the waves. At subcritical fields, isolated particles cannot be excited. Here, in this study, isolated particles can be excited directly by $E_p$ ($>E_c$) in each cycle. Similarly, it was found that as the stopping time of $E$ is short enough such that the motion of particles becomes continuous, waves can form as well due to the electrohydrodynamic flows (33). By contrast, the motion of wave particles in this study is intermittent, although the waves propagate with a persistent direction.

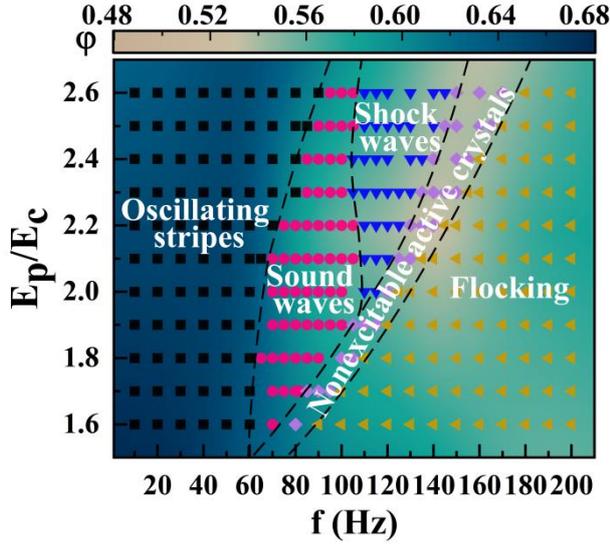

**Figure 6** Phase diagram. Five phases can be realized by tuning $E_p$ and $f$. At low frequencies, the displacements of particles in each cycle are far larger than the mean distance between particles, macroscopic stripes which splitting periodically with SWE forms. Upon the increase of frequency, as the displacements in each cycle become comparable with the mean distance between particles, no macroscopic density fluctuations form. The system condenses into active liquids in which local density fluctuations can be amplified and develop into sound waves. Further increasing the frequency, active crystals form in which density-dependent memory leads to the emergence of shock waves. However, as the displacements of particles in each cycle become smaller than the mean distance between particles, no density fluctuations emerge and shock waves are suppressed, resulting non-excitable active crystals. As the frequency is high enough such that the charge cannot be relaxed in each cycle, all particles remain a memory of the propulsion, flocking forms. In the condensed phases, however, the critical frequency for all particles to have a memory is lower than that for isolated particles. Scale bar: 150μm.

In summary, Quincke rollers driven by SWE exhibit tunable activity and density-dependent collective behaviors. When attractive interactions become significant with respect to activity, Quincke rollers aggregate and form active condensed phases. Distinct from normal condensed phases, the active condensed phases can be excited to form activity waves. In active liquids, sound waves emerge and propagate by splitting whereas in active crystals, shock waves emerge and propagate by density-enhanced collective memory. These two mechanisms have a common basis: density-dependent dynamic inhomogeneity. This inhomogeneity is attributed to the fact that the dynamics of Quincke rollers are sensitive to local density due to the dipolar interactions. Density fluctuations coupled with density-dependent dynamics break the local dynamic symmetry and induce local directed particle flows which results in a positive feedback loop to density fluctuations. As a result of the density-dependent dynamics, the waves observed in this study can only be excited and propagate in high density regions (Movie 12). This promises a pathway to prepare active systems with patterned density such that they can be selectively excited in space to form activity patterns. This is of great interests in fabricating dynamic structures for smart materials (25, 39-41).

**Materials and Methods**

**Experimental Setup**. Polystyrene spheres (Thermo Scientific G1000, polystyrene spheres) of diameter 9.9 $\mu$m are dispersed in a 0.12 mol/L AOT/hexadecane solution. AOT is dried at 80°C for 10 hours to remove the water completely. The suspension is then sealed between two indium-tin-oxide–coated glass slides which are separated by insulating spacers (Fig. 1 A, Top) with a thickness of 110±10$\mu$m thick. As a uniform electric field $\vec{E}$ is applied, the spheres are polarized. The dipole moment $\vec{P}$ is antiparallel to $\vec{E}$. As $\vec{P}$ is disturbed from its equilibrium state by fluctuations, a rotation torque $\vec{P} \times \vec{E}$ forms. To overcome the viscosity resistance of the fluid, the amplitude of $\vec{E}$ has to exceed a critical value $E_c$ to propel the particles to rotate and move (13, 30)

Dynamical processes are observed with a Nikon microscope (10×objective) and recorded at a rate of 1000 frames per second. The positions of particles are located and tracked with an accuracy of 100 nm (0.1 pix) using IDL codes(42).

**Velocity-velocity autocorrelation**

The velocity-velocity autocorrelation is calculated with the velocities of particles by:

$$C_v(\Delta t) = \sum_i \langle \frac{\vec{v}_i(t) \cdot \vec{v}_i(t+\Delta t)}{|\vec{v}_i(t)| \cdot |\vec{v}_i(t+\Delta t)|} \rangle_t$$

The results are averaged over time $t$ and all concerned particles $i$.

**Author contributions**
T.H.Z. designed the research. M.F.Z. and L.Z. performed experiments and contributed equally. T.H.Z. wrote the paper.

**Conflicts of interest**
There are no conflicts of interest to declare.

**Acknowledgements**
T.H.Z. acknowledges financial support of the National Natural Science Foundation of China (Grant 11974255). We thank Hugues Chaté and Xia-qing Shi for interesting discussions, and Hugues Chaté for a critical reading of the manuscript and valuable comments.

**References**

1. Ramaswamy S (2017) Active matter. *J. Stat. Mech. Theory Exp.* 2017(5):054002.

2. O'Byrne J, Kafri Y, Tailleur J, & van Wijland F (2022) Time irreversibility in active matter, from micro to macro. *Nat. Rev. Phys.* 4(3):167-183.
3. Vicsek T & Zafeiris A (2012) Collective motion. *Phys. Rep.* 517(3):71-140.
4. Needleman D & Dogic Z (2017) Active matter at the interface between materials science and cell biology. *Nat. Rev. Mater.* 2(9):17048.
5. Chaté H (2020) Dry Aligning Dilute Active Matter. *Annu. Rev. Condens. Matter Phys.* 11(1):189-212.
6. Sokolov A, Aranson IS, Kessler JO, & Goldstein RE (2007) Concentration dependence of the collective dynamics of swimming bacteria. *Phys. Rev. Lett.* 98(15).
7. Schaller V, Weber C, Semmrich C, Frey E, & Bausch AR (2010) Polar patterns of driven filaments. *NATURE* 467(7311):73-77.
8. Katz Y, Tunstrom K, Ioannou CC, Huepe C, & Couzin ID (2011) Inferring the structure and dynamics of interactions in schooling fish. *Proc. Natl. Acad. Sci. USA* 108(46):18720-18725.
9. Ballerini M, *et al.* (2008) Interaction ruling animal collective behavior depends on topological rather than metric distance: Evidence from a field study. *Proc. Natl. Acad. Sci. USA* 105(4):1232-1237.
10. Wang W, Lv X, Moran JL, Duan S, & Zhou C (2020) A practical guide to active colloids: choosing synthetic model systems for soft matter physics research. *Soft Matter* 16(16):3846-3868.
11. Zhang TH, Groenewold J, & Kegel WK (2009) Observation of a microcrystalline gel in colloids with competing interactions. *Phys. Chem. Chem. Phys.* 11(46):10827-10830.
12. Zhang TH & Liu XY (2014) Experimental modelling of single-particle dynamic processes in crystallization by controlled colloidal assembly. *Chem. Soc. Rev.* 43(7):2324-2347.
13. Bricard A, Caussin J-B, Desreumaux N, Dauchot O, & Bartolo D (2013) Emergence of macroscopic directed motion in populations of motile colloids. *Nature* 503(7474):95-98.
14. Bricard A*, et al.* (2015) Emergent vortices in populations of colloidal rollers. *Nat. Commun.* 6(1):7470.
15. Lu SQ, Zhang BY, Zhang ZC, Shi Y, & Zhang TH (2018) Pair aligning improved motility of Quincke rollers. *Soft Matter* 14(24):5092-5097.
16. Vicsek T, Czirók A, Ben-Jacob E, Cohen I, & Shochet O (1995) Novel Type of Phase Transition in a System of Self-Driven Particles. *Phys. Rev. Lett.* 75(6):1226-1229.
17. Zhang B, Karani H, Vlahovska PM, & Snezhko A (2021) Persistence length regulates emergent dynamics in active roller ensembles. *Soft Matter* 17(18):4818-4825.
18. Grégoire G & Chaté H (2004) Onset of Collective and Cohesive Motion. *Phys. Rev. Lett.* 92(2):025702.
19. Zhang B, Yuan H, Sokolov A, de la Cruz MO, & Snezhko A (2022) Polar state reversal in active fluids. *Nat. Phys.* 18(2):154-159.
20. Henkes S, Fily Y, & Marchetti MC (2011) Active jamming: Self-propelled soft particles at high density. *Phys. Rev. E* 84(4):040301.
21. Geyer D, Martin D, Tailleur J, & Bartolo D (2019) Freezing a Flock: Motility-Induced Phase Separation in Polar Active Liquids. *Phys. Rev. X* 9(3):031043.
22. Menzel AM & Löwen H (2013) Traveling and Resting Crystals in Active Systems. *Phys. Rev. Lett.* 110(5):055702.
23. Bialké J, Speck T, & Löwen H (2012) Crystallization in a Dense Suspension of Self-Propelled Particles. *Phys. Rev. Lett.* 108(16):168301.
24. Briand G, Schindler M, & Dauchot O (2018) Spontaneously Flowing Crystal of Self-Propelled Particles. *Phys. Rev. Lett.* 120(20):208001.
25. Zhang J, Luijten E, Grzybowski BA, & Granick S (2017) Active colloids with collective mobility status and research opportunities. *Chem. Soc. Rev.* 46(18):5551-5569.
26. Li Z, Fan Q, & Yin Y (2022) Colloidal Self-Assembly Approaches to Smart Nanostructured Materials. *Chem. Rev.* 122(5):4976-5067.
27. Leanza S, Wu S, Sun X, Qi HJ, & Zhao RR (2024) Active Materials for Functional Origami. *Adv. Mater.* 36(9):2302066.
28. Baconnier P*, et al.* (2022) Selective and collective actuation in active solids. *Nat. Phys.* 18(10):1234-1239.
29. Caprini L, Marini Bettolo Marconi U, Puglisi A, & Löwen H (2023) Entropons as collective excitations in active solids. *J. Chem. Phys.* 159(4).
30. Jones TB (1984) Quincke Rotation of Spheres. *IEEE Transactions on Industry Applications* IA-20(4):845-849.
31. Liu ZT*, et al.* (2021) Activity waves and freestanding vortices in populations of subcritical Quincke rollers. *Proc. Natl. Acad. Sci. USA* 118(40):e2104724118.
32. Das D & Saintillan D (2013) Electrohydrodynamic interaction of spherical particles under Quincke rotation. *Phys. Rev. E* 87(4):043014.
33. Zhang B, Glatz A, Aranson IS, & Snezhko A (2023) Spontaneous shock waves in pulse-stimulated flocks of Quincke rollers. *Nat. Commun.* 14(1):7050.
34. Yang Y, Zhang ZC, Qi F, & Zhang TH (2022) Emergence of Self-dual Patterns in Active Colloids with Periodical Feedback to Local Density. *arXiv:2204.07717*.
35. Zhang B, Snezhko A, & Sokolov A (2022) Guiding Self-Assembly of Active Colloids by Temporal Modulation of Activity. *Phys. Rev. Lett.* 128(1):018004.
36. Zhang ZC (2019) Dynamic Self-Assembly in Self-propelled Colloids. Master (Soochow University).
37. Yao LD (2021) Emergence of Space-Time Patterns in Self-propelled Colloidal Systems. Master (Soochow University).
38. Yang Y, Meng Fei Z, Lailai Z, & Tian Hui Z (2023) Tunable Memory and Activity of Quincke Particles in Micellar Fluid. *Chin. Phys. Lett.* 40(12):126401.
39. Wang W, Duan W, Ahmed S, Sen A, & Mallouk TE (2015) From One to Many: Dynamic Assembly and Collective Behavior of Self-Propelled Colloidal Motors. *Acc. Chem. Res.* 48(7):1938-1946.
40. Grzybowski BA, Fitzner K, Paczesny J, & Granick S (2017) From dynamic self-assembly to networked chemical systems. *Chem. Soc. Rev.* 46(18):5647-5678.
41. Merindol R & Walther A (2017) Materials learning from life: concepts for active, adaptive and autonomous molecular systems. *Chem. Soc. Rev.* 46(18):5588-5619.
42. Crocker JC & Grier DG (1996) Methods of Digital Video Microscopy for Colloidal Studies. *J. Colloid Interface Sci.* 179(1):298-310.

# Activity Waves in Condensed Phases of Quincke Rollers

Meng Fei Zhang,‡[a]   Bao Ying Fan,‡[a]   Zeng Tao Liu,[a]   Tian Hui Zhang[a]

[a] *Center for Soft Condensed Matter Physics and Interdisciplinary Research & School of Physical Science and Technology, Soochow University, Suzhou, 215006, P. R. China*
[b] *Service de Physique de l'Etat Condensé, Commissariat à l'Energie Atomique (CEA), CNRS, Université Paris-Saclay, CEA-Saclay, 91191 Gif-sur-Yvette, France;*
[c] *Computational Science Research Center, Beijing 100193, China*

# Supplementary Information

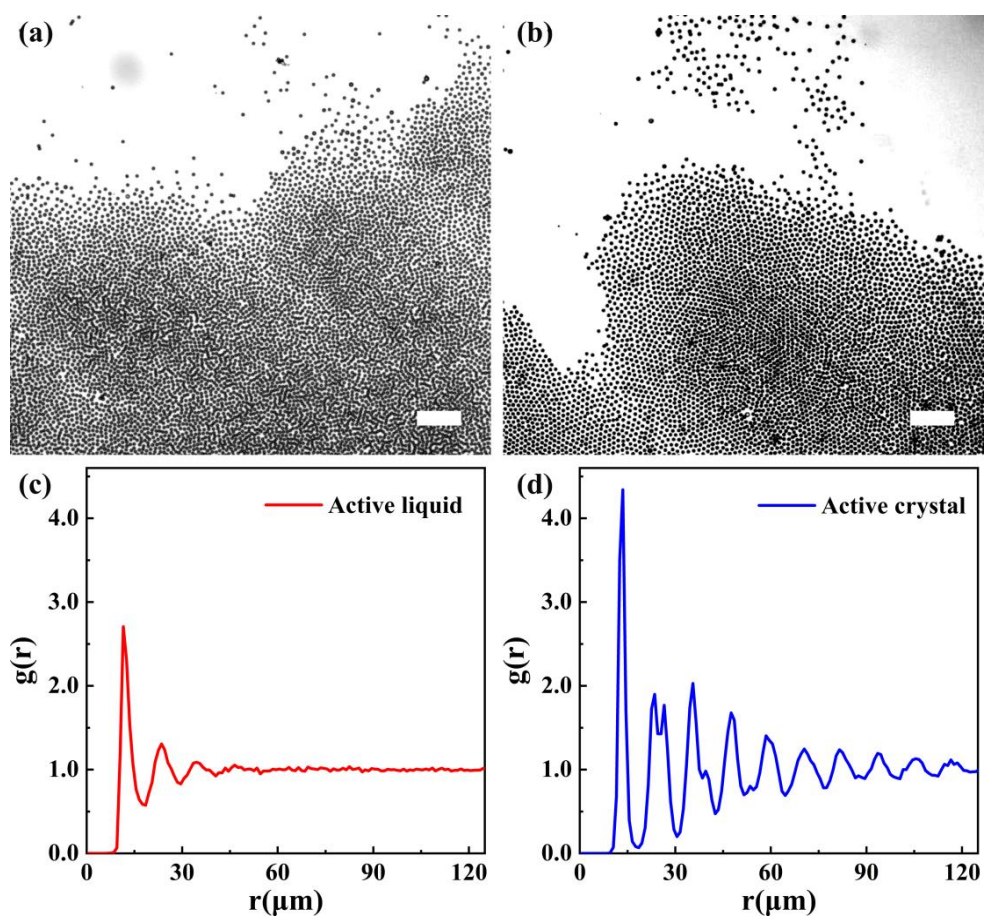

**SFigure 1** As the system condensed, an empty region forms between the cell boundary and the condensed phase. a: Active liquid. $E_p = 2.5 E_c$ and $f = 100$Hz.   b: Active crystal. $E_p = 2.5 E_c$ and $= 120$Hz . c: In active liquids, the pair distribution function exhibits no long-ranged correlation. d: The pair distribution function in active crystals exhibit sharp peaks where the second peak consists of two sub-peaks which are the typical sign of a hexagonal symmetry.  Scale bar in a and b: 150μm.

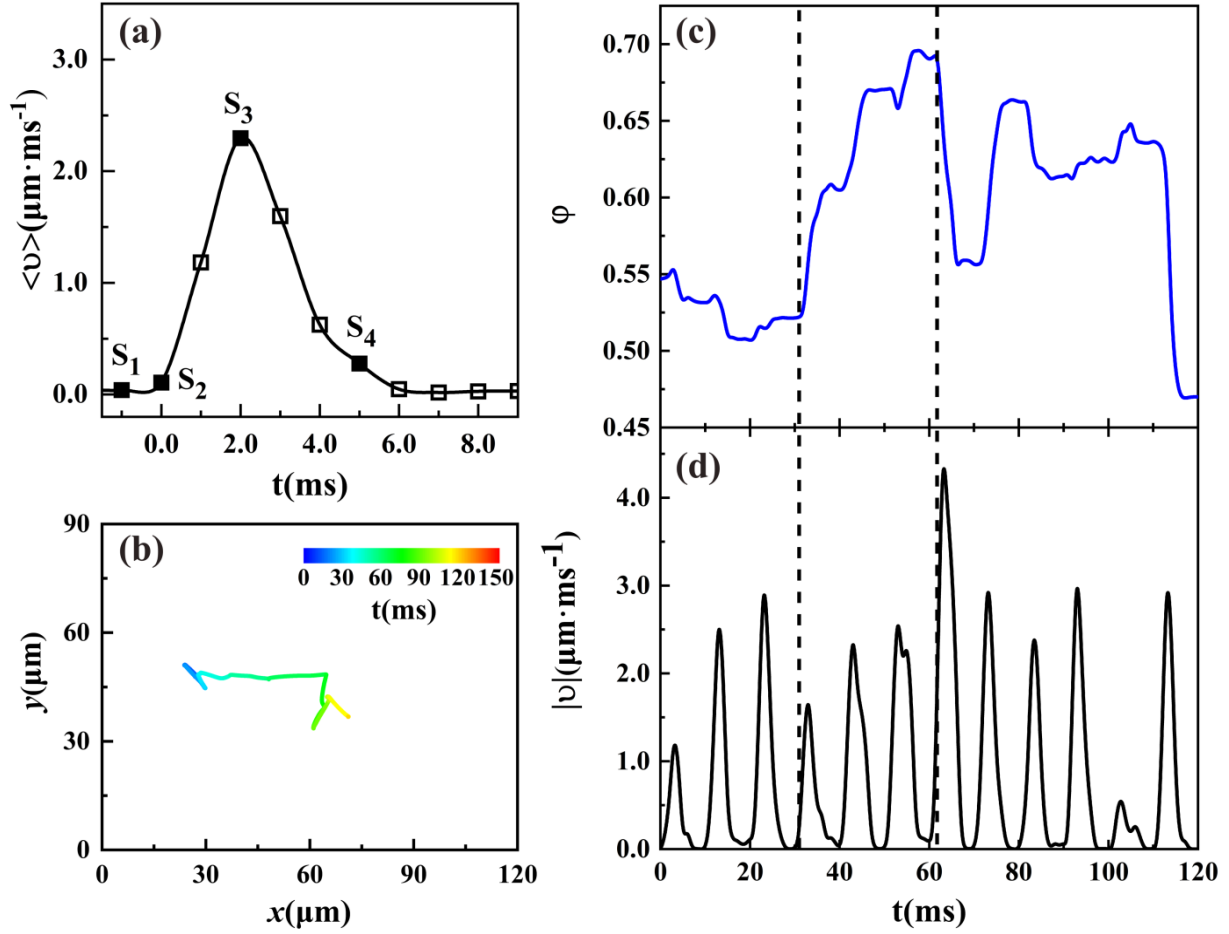

**SFigure 2** a. Mean speed of particles in active liquid as a function of time in one cycle when no wave passes. $E_p = 2.5 E_c$ and $f = 100$Hz. The four points $S_1, S_2, S_3, S_4$ are the corresponding times of snapshots shown in Fig.3a-d where one wave passes. b. A typical trajectory of particles in wave. As particles are included in the wave, their motions become persistently directed in the following cycles. Before and after the wave passes, their motions are randomly driven in each cycle. The directed displacements of wave particles are around 30μm which is much smaller than the width of wave. c. The local density observed by one wave particle experiences a sharp increase as it is included in the wave front, and a sharp drop as it is separated from the wave by splitting. d. The maximum speed of wave particles is much higher than the peak value measured in the absence of wave. Nevertheless, the speed of wave particles varies not only with time but also with their positions in the wave. Therefore, the speed curves of wave particles as a function of time are characterized by different shapes in different cycles. In active fluids, a wave passes one particle by three cycles in most cases.

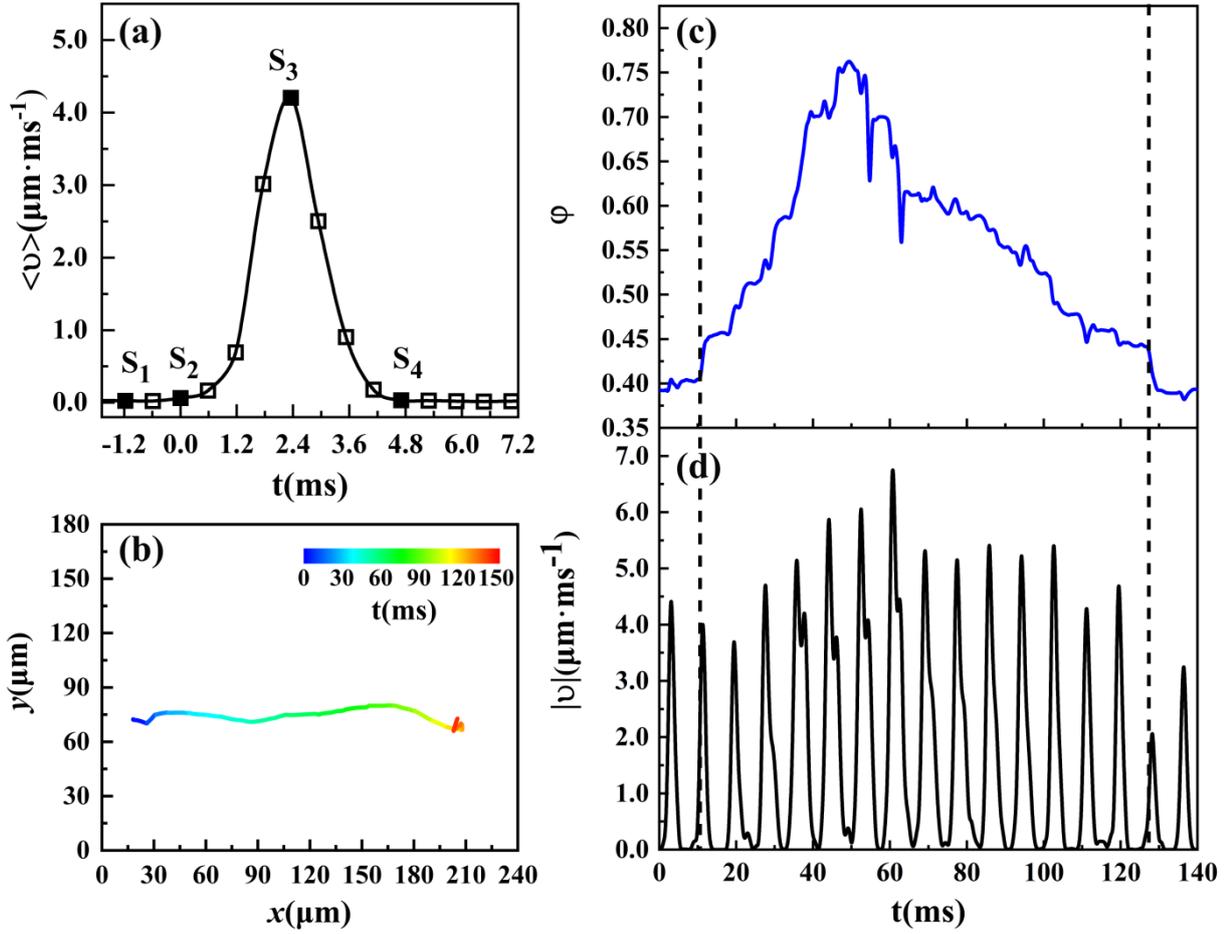

**SFigure 3** Mean speed of particles in active crystals as a function of time in one cycle when no wave passes. $E_p = 2.5E_c$ and $f = 120$Hz. The four points marked by $S_1, S_2, S_3, S_4$ are the corresponding times of snapshots shown in Fig.4a-d where one wave passes. b. A typical trajectory of particles in wave. The displacements of wave particles can be as high as 180μm which is also much smaller than the width of wave. c. The local density observed by one wave particle. As one particle is involved in the wave front, the density increases sharply above 0.45. As it drops off from the wave tail, the density drops down to 0.40. d. The maximum speed of wave particles is much higher than the peak value measured in the absence of wave. The speed of wave particles varies with time and their positions in the wave. In active crystals, a wave passes one particle by more than ten cycles. However, both the width of wave and the number of cycles experienced by wave particles vary from case to case.

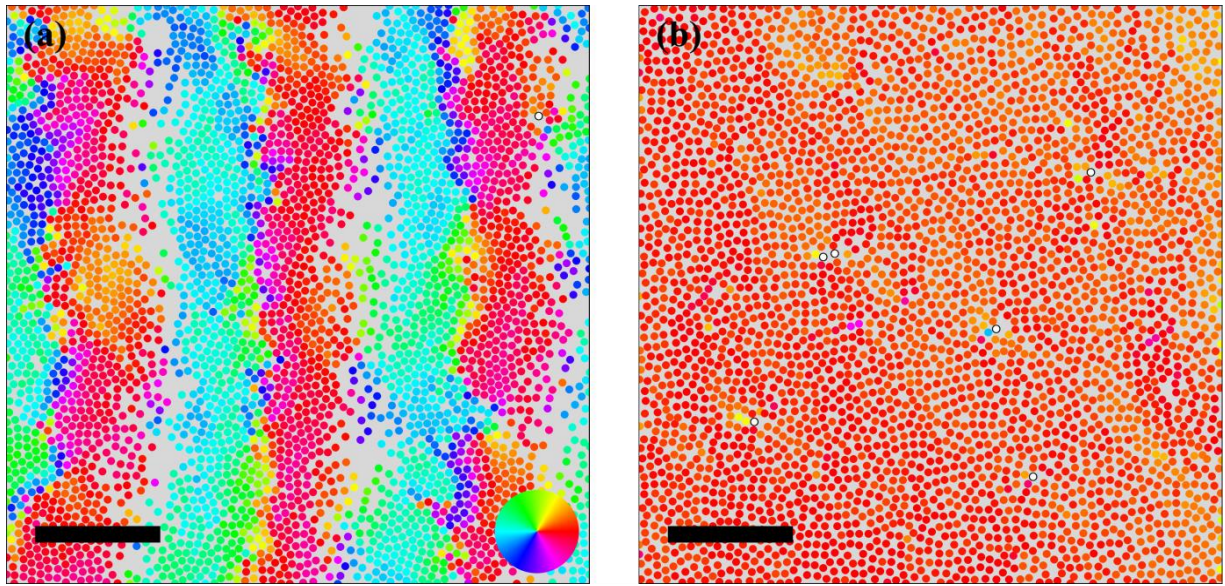

**SFigure 4** Snapshots of phases. a. Oscillating stripes. b. Flocking in which particles move continuously Particles are colored according to their directions of velocity. Scale bar: 150μm.